\documentclass{svmult}
\usepackage{amssymb,amscd}

\newcommand{\beq}{\begin{equation}}
\newcommand{\eeq}{\end{equation}}
\newcommand{\bea}{\begin{eqnarray}}
\newcommand{\eea}{\end{eqnarray}}
\newcommand\la{{\lambda}}
\newcommand\ka{{\kappa}}
\newcommand\al{{\alpha}}
\newcommand\be{{\beta}}
\newcommand\ep{{\epsilon}}
\newcommand\si{{\sigma}}

\newcommand\lu{{\bf U}}
\newcommand\lv{{\bf V}}
\newcommand\sur{{\mathcal S}}
\newcommand\hps{{\hat{\psi}}}
\newcommand\bps{{\overline{\psi}}}
\newcommand\uc{{\underline{\mathrm{c}}}}
\newcommand\ul{{\underline{\ell}}}
\newcommand\ug{{\underline{\mathrm{g}}}}
\newcommand\uom{{\underline{\omega}}}
\newcommand\om{{\omega}}

\begin{document}

\title*{Painlev\'{e} tests, singularity structure and integrability}
\author{Andrew N.W. Hone}
\institute{Institute of Mathematics \& Statistics, University of
Kent,  Canterbury CT2 7NF, UK \texttt{anwh@kent.ac.uk }}
\maketitle

\begin{abstract}
After a brief introduction to the Painlev\'{e} property for
ordinary differential equations, we present a concise
review of the various methods
of singularity analysis which are commonly referred to as
Painlev\'{e} tests. The tests are applied to
several different examples, and the connection between
singularity structure and integrability for ordinary and
partial differential equation is discussed. 
\end{abstract}

\index{Painlev\'e test} 
\section{Introduction} 

The connection between the integrability of differential equations 
and the singularity structure of their solutions was first 
discovered in the pioneering work of Kowalevski \cite{kow}, 
who considered the equations for the motion under gravity of a rigid 
body about a fixed point, namely  
\beq 
\begin{array}{ccl} 
d\ul & = & \ul \bf{\times} \uom + \uc \bf{\times} \ug\, , \\ 
\overline{dt} && \\ 
d\ug & = & \ug \bf{\times} \uom\, ; \qquad \qquad \ul =\bf{I}\, \uom\, . \\ 
\overline{dt} && 
\end{array} 
\label{kt} 
\eeq 
In the above, $\ul$ and $\uom$ are respectively the angular momentum 
and angular velocity of the body, $\ug$ is the gravity vector with 
respect to 
a moving frame, and the centre of mass vector $\uc$ and 
inertia tensor $\bf{I}$ are both constant. 
The remarkable insight of Kowalevski was that the system of 
equations (\ref{kt}) could be solved explicitly whenever the 
dependent variables $\ul$ and $\ug$ are {\it meromorphic} 
functions of time $t$ extended to the complex plane, 
$t\in \mathbb{C}$.  By requiring that the solutions should 
admit Laurent expansions around singular points, she 
found constraints on the constants $\uc$ and ${\bf I} = 
\mathrm{diag}(I_1,I_2,I_3)$ (diagonalized in a suitable frame). 
Her method isolated the two solvable cases previously known to 
Euler ($\uc =0$) and Lagrange ($I_1=I_2$ with $\uc$ defining 
the axis of symmetry), as well as a new case having 
principal moments of inertia $I_1=I_2=2I_3$ and $\uc$ 
perpendicular to the axis of symmetry. The latter case is now 
known as the Kowalevski top, 
\index{Kowalevski top} 
and Kowalevski was further 
able to integrate it explicitly in terms of theta-functions 
associated with a hyperelliptic curve of genus 2, 
thereby proving directly that the solutions are 
meromorphic functions of $t$. A modern discussion can be found in 
\cite{babelonetal} or \cite{mark}, for instance. 

An important feature of the equations (\ref{kt}) from the 
point of view of singularity analysis is that they are 
{\it nonlinear}. For a linear differential equation 
$$ 
\frac{d^n y}{dz^n}+a_{n-1}(z)\frac{d^{n-1} y}{dz^{n-1}} 
+\ldots +a_1(z)\frac{d y}{dz} +a_0(z)y=0, 
$$ 
of arbitrary order $n$ it is well known \cite{hille, ince} 
that the general solution can have only {\it fixed} singularities 
at the points in the complex $z$-plane where the 
coefficient functions $a_j(z)$ are singular. However, for nonlinear 
differential  equations, 
as well as the fixed singularities which are 
determined by the equation itself, the 
solutions can have {\it movable} 
singularities which vary with the initial conditions. For example, 
the first order nonlinear differential equation 
$$ 
\frac{d y}{dz} +y^2=0 
$$ 
has the general  solution 
$$ 
y=\frac{1}{z-z_0}, \qquad z_0 \quad \mathrm{arbitrary}, 
$$ 
with a movable simple pole at $z=z_0$. If the initial data $y=y_0$ 
is specified at the point $z=0$, then the position of the simple 
pole varies according to 
$$ 
z_0=-\frac{1}{y_0}. 
$$ 
 
The classification of ordinary differential equations (ODEs) 
in terms of their singularity structure was initiated 
in the work of Painlev\'{e} \cite{painleve}. The main property 
that Painlev\'e sought for ODEs was that their solutions should 
be single-valued around movable singular points. Nowadays 
this property is usually formulated thus: 

\index{Painlev\'e property} 
 
\vspace{.1in} 
\noindent {\bf Definition 1.1. The Painlev\'e property for ODEs: } 
{\it An ODE has the Painlev\'e property if all movable 
singularities of all solutions are poles.} 
\vspace{.05in} 
 
\noindent Painlev\'e proved that for first order ODEs 
of the general form 
$$ 
y'=\frac{\mathcal{P}(y,z)}{\mathcal{Q}(y,z)}, 
$$ 
where $\mathcal{P}$ and $\mathcal{Q}$ are polynomial 
functions of $y$ and analytic functions of $z$ 
(and the prime $'$ denotes $d/dz$), then 
the only movable singularities that can arise are 
poles and algebraic branch points. The latter are 
excluded by Definition 1.1, and he further showed that 
the most general first order equation with the Painlev\'e 
property is the 
\index{Riccati equation} 
Riccati equation 
$$ 
y'=a_2(z)y^2+a_1(z)y+a_0(z), 
$$ 
where the coefficients $a_0$, $a_1$, $a_2$ are 
analytic functions of $z$. 
 
For second order ODEs, life is more complicated 
because movable essential singularities can arise (see e.g. chapter 3 in 
\cite{abfo} for an example). 
Painlev\'{e} initiated the classification of 
second order ODEs of the form 
\begin{equation} 
y''=\mathcal{F} (y',y,z). 
\label{ptype} 
\end{equation} 
with $\mathcal{F}$ being a rational function of $y'$ and 
$y$, and analytic in $z$. Painlev\'{e} and his contemporaries 
succeeded in classifying  all ODEs of the type (\ref{ptype}) which 
fulfill the requirements of Definition 1.1. 
The complete result is in the form of a list of 
approximately fifty representative equations, unique up to 
M\"obius transformations, which are detailed in 
chapter 14 of Ince's book \cite{ince}. 
It was found that (after suitable  
changes of variables) all of these ODEs have general 
solutions in terms of 
classical special functions (defined by linear equations) or elliptic 
functions, except for six special 
equations which are now known as  Painlev\'{e} I-VI (or just PI-VI). 
 
As an example, consider the second order ODE 
\beq 
\label{ellip} 
y''=6y^2-\frac{1}{2}g_2. 
\eeq 
This can be immediately integrated once, 
because the equation is autonomous (the right hand side is 
independent of $z$), which yields 
\beq 
\label{wp} 
(y')^2=4y^3-g_2y-g_3, 
\eeq 
with $g_3$ being an integration constant. 
The general solution of the first order ODE (\ref{wp}) is given by the 
Weierstrass elliptic function, 
\beq 
y=\wp (z-z_0;g_2,g_3) 
\label{ellsoln} 
\eeq 
with the constants $g_2$, $g_3$ 
being the invariants. The solution 
(\ref{ellsoln}) has infinitely many movable double poles, at 
$z=z_0$ and at all 
congruent points 
$z=z_0+ 2m \om_1+2n \om_2 \in \mathbb{C}$ 
for $(m,n) \in \mathbb{Z}^2$ 
on the period lattice defined by the half-periods $\om_1,\om_2$. 
(For an introduction to Weierstrass elliptic functions see   
chapter 20 in \cite{ww} or chapter VI in \cite{silverman}.)    
\index{elliptic functions} 
We shall return to the equation 
(\ref{ellip}) in the next section. 
 
The first of the {\it Painlev\'{e} equations} is PI, 
\index{First Painlev\'e equation} 
which is a 
non-autonomous version of (\ref{ellip}) given by 
\beq 
\label{p1} 
y''=6y^{2}+z. 
\eeq 
Its general solution is a meromorphic function of 
$z$, and the solution of PI (or sometimes the equation itself) 
may be referred to as a Painlev\'{e} transcendent, since it 
essentially 
defines a new transcendental function. 
The other equations PII-PVI also contain parameters; for 
example the second Painlev\'e equation (PII) 
\index{Second Painlev\'e equation} 
is 
\beq 
\label{p2} 
y''=2y^{3}+zy+\alpha 
\eeq 
where $\al$ is a constant parameter. The general solution of each of 
the  Painlev\'e equations cannot be expressed in terms of 
elliptic functions or other classical special functions \cite{iksy}, 
although for special 
parameter values they can be solved in this way; e.g. 
when $\al$ is an integer, equation 
(\ref{p2}) has particular solutions given by rational functions 
of $z$, and it has solutions in terms of Airy functions  
for half-integer values of $\al$. 
 
An important early result was the connection of 
PVI with the isomonodromic deformation of an associated 
linear system \cite{fuchs}. 
After the work of Painlev\'e and his colleagues around the 
turn of the last century, the Painlev\'e equations 
were probably only of interest to experts on differential 
equations. 
However, in the latter half of the 20th 
century the Painlev\'e transcendents enjoyed 
something of a renaissance when it was discovered 
that they gave exact formulae for correlation functions 
in solvable models of statistical mechanics \cite{mccoy}, 
quantum field theory \cite{iiks} and random matrix models 
\cite{douglas, jmms}, 
and also arose as symmetry reductions of partial differential 
equations (PDEs) integrable by the inverse scattering transform 
(see \cite{ars} and section 3 below). 
The link with integrable PDEs and linear Lax pairs 
established the exact solution of the Painlev\'e equations 
by the isomonodromy method \cite{fla}. 
More recently a weakened version of the Painlev\'e 
property has been used to find exact metrics for 
relativistic fluids \cite{rod}. 
With this wide variety of physical applications, 
the Painlev\'e transcendents have acquired the status 
of nonlinear special functions 
(see the review and references in chapter 7 of \cite{abcla}). 
 
The continuation of Painlev\'e's classification programme 
to higher order equations becomes increasingly difficult 
as the order increases. Even at third order a new phenomenon 
can be encountered, in the form of a movable natural barrier or boundary 
beyond which the solution cannot be analytically continued; this 
occurs in  Chazy's equation 
\index{Chazy equation} 
\beq 
y'''=2yy''-3(y')^2. 
\label{chazy} 
\eeq 
A variety of results for third or higher order equations have been obtained 
by Chazy \cite{chazy}, Gambier, Bureau, and most recently by 
Cosgrove; see \cite{cosgrove} and references therein. 
Chazy's equation (\ref{chazy}) has some higher order relatives known as 
Darboux-Halphen systems, which have a very 
complicated singularity structure, and occur as reductions of the 
integrable self-dual Yang-Mills equations (see the contribution 
of Ablowitz {\it et al} in \cite{cargese}).

It should be  clear from the above that the Painlev\'e 
property has a very deep connection with the concept 
of integrability. This connection is by no means straighforward, 
and continues to be the subject of active research \cite{cargese}. 
In the rest of this brief review article, we will introduce the basic 
techniques for testing the singularity structure of 
differential equations (both ODEs and PDEs), which 
are often referred to collectively as {\it Painlev\'e analysis}. 
The basic method for testing ODEs by expansions in 
Laurent series is treated in section 2.  
This method 
should probably  
be referred to as the {\it Kowalevski-Painlev\'e test} to  
honour 
both pioneers of the subject, but most commonly only 
Painlev\'e is mentioned in  
this context. 
Section 3 describes the conjecture of Ablowitz, Ramani and Segur \cite{ars} 
on the connection between integrable PDEs and Painlev\'e-type ODEs, 
and how 
this can be used as an integrability test for PDEs. In the fourth 
section we explain how the preceding analysis can 
be bypassed by a direct consideration of the singularity structure 
of a PDE, by using the method of Weiss, Tabor and 
Carnevale \cite{wtc}. This is followed in  
section 5 
by associated truncation techniques related to B\"acklund  
transformations, 
Lax pairs and Hirota bilinear equations for integrable  
systems of PDEs.  
In section 6 we highlight the  
limitations of the 
Painlev\'e property as a criterion for integrability,  
in the context 
of integrable systems with movable algebraic branching 
and the weak  
Painlev\'e property \cite{weak}. 
\index{Weak Painlev\'e property}  
In the  
final section we give our outlook on 
methods of singularity analysis for  
differential 
equations, and mention how some of these methods 
and  concepts  
have been extended to the discrete context of maps or difference equations.

\section{Painlev\'e analysis for ODEs} 

Consider an ODE for a dependent variable $y(z)$, which may be a single scalar, 
or a vector quantity. 
If the ODE has the 
Painlev\'e property then its solutions must have local Laurent expansions 
around movable singularities at $z=z_0$, where $z_0$ is arbitrary. 
However, if branching occurs then this can be detected by 
local singularity analysis. 
\index{Painlev\'e test}
The basic Painlev\'e test for ODEs consists of the following steps: 
\begin{itemize} 
\item {\bf Step 1:} Identify all possible {\it dominant balances}, 
i.e. all singularities of form $y\sim c_0 \, (z-z_0)^\mu$. 
\item {\bf Step 2:} If all exponents $\mu$ are integers, find the 
{\it resonances} where arbitrary constants can appear. 
\item {\bf Step 3:}  If all resonances are integers, check the 
{\it resonance conditions} in each Laurent expansion. 
\item {\bf Conclusion:} If no obstruction is found in steps 1-3 for 
every dominant balance then 
the Painlev\'e test is satisfied. 
\end{itemize} 
Note that the exponents $\mu$ and leading coefficients $c_0$ must have 
as many components as the vector $y$, and if the ODE is 
polynomial then at least one of the exponents must be a {\it negative} 
integer for a leading order pole-type singularity. Rather than give formal 
definitions of the terms introduced  in steps 1-3 above (which can be 
found in \cite{chang} and elsewhere in the references), 
we would like to illustrate them with a couple of examples. 
 
First of all we describe the Painlev\'e test applied to the equation 
(\ref{ellip}), in which case $y$ is just a scalar. Applying step 1 we 
look for leading order behaviour which produces a  singularity in the ODE, 
so we require $y\sim c_0 \, (z-z_0)^\mu$ and $\mu$  must be a negative 
integer for a movable pole with no branching. This gives 
immediately 
\beq 
y\sim \frac{1}{(z-z_0)^2} 
\label{leadel} 
\eeq 
as the only possible dominant balance. Note that we could have also 
obtained this balance by assuming that $y$ blows up as $z\to z_0$, 
and  then (since $g_2$ is constant) $y^2>>g_2$  on the 
right hand side of the ODE, so the $y^2$ term must balance with the 
left hand side of 
(\ref{ellip}), giving 
\beq 
\label{bal} 
y''\sim 6y^2, \qquad \mathrm{as}\quad z\to z_0. 
\eeq 
We can multiply by $y'$ on both sides of (\ref{bal}) and integrate to 
find 
\beq 
\label{bal2} 
\frac{1}{2}(y')^2\sim 2y^3,  \qquad \mathrm{as}\quad z\to z_0 
\eeq 
(throwing away the integration constant, which is strictly dominated 
by the other terms), and after taking a square root 
in (\ref{bal2}) and integrating we find (\ref{leadel}). 
 
We now seek a solution of (\ref{ellip}) given 
locally by a Laurent expansion around a double pole at $z=z_0$, in 
the form 
\beq 
y=\sum_{j=0}^\infty c_j\, (z-z_0)^{j-2}, \qquad c_0=1, 
\label{elexpa} 
\eeq 
where the value of $c_0$ has been fixed as in (\ref{leadel}). 
We  wish to determine the {\it resonances}, which 
are the positions in the Laurent series (\ref{elexpa}) where 
arbitrary coefficients $c_j$ can appear. Since the ODE (\ref{ellip}) 
is second order, there must be two arbitrary constants in 
a local representation of the general solution: $z_0$, the arbitrary 
position of the movable pole, and one other. 
To apply step 2 of the Painlev\'e test we take a perturbation 
of the leading order with small parameter $\ep$, in the 
form 
\beq 
y\sim (z-z_0)^{-2} (1+\ep (z-z_0)^r). 
\label{pertep} 
\eeq 
To first order in $\ep$  we have 
$$ 
y^2 \sim (z-z_0)^{-4} (1+2\ep (z-z_0)^r), \qquad 
y'' \sim (z-z_0)^{-4} (6+\ep (r-2)(r-3) (z-z_0)^r). 
$$ 
Thus when we substitute the perturbation (\ref{pertep}) 
into the dominant terms (\ref{bal}) 
and retain only first order terms in $\ep$ we find 
$$ 
y''-6y^2\sim \ep \Big( (r-2)(r-3)-12 \Big) (z-z_0)^{r-4}=0. 
$$ 
Since the perturbation $\ep$ is arbitrary, corresponding 
to the first appearance of a new arbitrary constant in the 
Laurent expansion (\ref{elexpa}), the expression in 
large brackets must vanish, giving the resonance polynomial 
$$ 
r^2-5r-6=0, \qquad \mathrm{whence} \quad r=-1 \quad \mathrm{or} \quad r=6. 
$$ 
The first resonance at $r=-1$ must {\it always} be present in 
any expansion around a movable singularity, since it corresponds to 
the arbitrariness of $z_0$. The second resonance value 
$r=6$ indicates that the coefficient $c_6$ should be arbitrary. 
 
In order to complete the Painlev\'e test, we must now substitute in 
the full Laurent expansion and check that it is consistent up 
to the coefficient $c_6$. In this case we find that the expansion is 
precisely 
\beq 
y=\frac{1}{(z-z_0)^2}+\frac{1}{20}g_2(z-z_0)^2 
+\frac{1}{28}g_3 (z-z_0)^4+\ldots , 
\label{wpser} 
\eeq 
so that $c_6=g_3/28$ is the arbitrary constant 
that appears after integrating (\ref{ellip}) to obtain 
(\ref{wp}). In fact only even powers of $(z-z_0)$ occur in this 
expansion, since the Weierstrass function (\ref{ellsoln}) is an 
even function of its argument. The higher coefficients 
in (\ref{wpser}) can be found recursively in terms of the 
invariants $g_2$, $g_3$. (Up to overall  
multiples 
these coefficients are the  
Eisenstein series 
associated to the corresponding  
elliptic curve \cite{silverman}.) 
The pole position $z_0$ does not appear 
in the coefficients because the ODE (\ref{ellip}) is 
autonomous. 
 
Here we should point out that 
passing the basic Painlev\'e test is only a {\it necessary} 
condition for an ODE to have the Painlev\'e property. 
Proving the Painlev\'e property 
requires showing that the local  
Laurent expansions can be 
analytically continued globally to a  
single-valued function 
(or one with only fixed branched points), 
in the absence of movable  
essential singularities. For the 
ODE (\ref{ellip}) this follows from the fact that the 
general solution (\ref{ellsoln}) is given by a 
Weierstrass elliptic function, which is meromorphic (for a 
proof see e.g. \cite{silverman, ww}). Painlev\'e's proof that the first 
Painlev\'e transcendent (\ref{p1}) is free from 
movable essential singularities is outlined by Ince in 
chapter 14 of \cite{ince}, but the proof is unclear and this has 
prompted recent  
efforts to find a more straightforward 
approach \cite{laine, jk, steinmetz}. 
 
Having seen an example where the Painlev\'e test is passed, we now 
move on to an example for which it fails, by considering the 
following coupled second order system: 
 \beq 
y_1''= 2y_1^2 -12y_2, 
\qquad y_2''=2y_1y_2. \label{sys4} \eeq 
In \cite{davmat} this system is associated to an interaction of four particles 
moving in a plane, subject to velocity-dependent forces, 
and in that context it is essential that both $y_1(z)$,  $y_2(z)$ (denoted 
$c_2(\tau)$, $c_4(\tau)$ in the original reference) and the independent 
variable $z$ should be 
{\it complex}. 
To find the dominant 
balances, we look for leading order singular behaviour of the form 
\begin{equation} 
y_1 \sim a Z^\mu , \qquad y_2\sim b Z^\nu , \label{asym} 
\end{equation} 
corresponding to a singularity in the solution at $Z=z-z_0 =0$ for 
at least one of $\mu$, $\nu$ negative. Because the system (\ref{sys4}) 
is {\it autonomous}, we can expand in the variable $Z$, since the position 
$z_0$ of the 
movable singularity will not appear in the coefficients of local expansions 
around $z=z_0$. 
 
There are three possible dominant  balances for the system 
(\ref{sys4}), namely 
$$ 
\begin{array}{rll} 
 (i) & 
\quad y_1\sim 3Z^{-2}, \quad y_2\sim bZ^{-2}, \qquad 
 & b\quad \mathrm{arbitrary} ; \\ 
(ii) & \quad y_1\sim 3Z^{-2}, 
 \quad y_2\sim bZ^3, 
 & b\quad \mathrm{arbitrary} ; \\ 
(iii) & \quad y_1\sim 10Z ^{-2}, \,\, y_2\sim\frac{35}{3}Z 
^{-4}. & 
\end{array} 
$$ 
Other possible power law behaviour around $Z =0$ corresponds 
to $\mu$, $\nu$ both non-negative integers and leads to Taylor 
series expansions, which are not relevant to our analysis of 
singular points. 
 
The second step in applying the Painlev\'{e}  test is to find 
the resonances. 
For the system (\ref{sys4}) to 
possess the Painlev\'{e} 
property we require that 
all resonances for all dominant balances must be integers, and at 
least one balance must have one resonance value of $-1$ with the 
rest being non-negative integers, in which case this is a 
{\it principal balance} for which the Laurent expansion should provide 
a local representation of the general solution. To find the 
resonance numbers $r$ we substitute 
$$ 
y_1\sim aZ ^\mu (1+\delta Z ^r ), \qquad y_2\sim bZ ^\nu 
 (1+\epsilon Z 
^r) 
$$ 
into the dominant terms of the system (\ref{sys4}) for each of the 
balances $(i)-(iii)$, and take only the terms linear in $\delta$ 
and $\epsilon$. This yields a pair of homogeneous linear 
equations for $\delta$, $\epsilon$ (which correspond to the 
arbitrary coefficients appearing at the resonances). The 
determinant of this $2\times 2$ system must vanish, which gives 
in each case a fourth order polynomial in $r$. 
 
\noindent {\bf Principal balance (i):} It turns out that the 
balance $(i)$ is the only principal balance, with resonances 
$$ 
(i) \qquad r=-1,0,5,6. 
$$ 
As mentioned before, the resonance $-1$ is always present, 
since it corresponds to the 
arbitrary position $z_0$ of the pole, while $r=0$ comes from 
the  arbitrary constant $b$ in the leading order term of the 
expansion for $y_2$; the other two values arise from arbitrary 
coefficients higher up in the series for $y_1$, $y_2$, so that 
altogether there should be four arbitrary constants appearing in 
these Laurent series. However, for step 3 of the test 
we also  require that all resonance conditions 
hold: so far we have only found the orders in the series where 
arbitrary constants may appear, but it is necessary to check that 
all other terms vanish at this order when the series are 
substituted into the equations. Taking \beq y_1\sim L_1(Z 
):=\sum_{j=-2} ^\infty k_{1,j}Z ^j, 
 \qquad 
y_2\sim L_2(Z ):=\sum_{j=-2} ^\infty k_{2,j}Z ^j 
\label{laurent} \eeq in the each of the equations (\ref{sys4}) we 
know already that the leading order terms require 
$$ 
k_{1,-2}=3,  \qquad k_{2,-2}=b\,\, (\mathrm{arbitrary}) , 
$$ 
giving the resonant term at $r=0$ in the expansion for $y_2$, 
while at subsequent orders we find 
$$ 
k_{1,-1}=0=k_{2,-1};\quad k_{1,0}=b,\, k_{2,0}= -b^2/3; \quad 
k_{1,1}=0=k_{2,1}. 
$$ 
At the next orders we further obtain 
$$ 
k_{1,2}=-3 b^2/5, \, k_{2,2}=7 b^3/15; \quad  k_{1,3}=0, \, 
k_{2,3}\quad  \mathrm{arbitrary} , 
$$ 
so that the resonance condition at $r=5$ corresponding to 
$k_{2,3}$ is satisfied. However, at the next order in the first 
equation of the system (\ref{sys4}), at the first appearance of 
the resonance coefficient $k_{1,4}$,  we find the additional 
relation 
$$ 
k_{2,2}= -b^3/5, 
$$ 
which means that the resonance condition is not satisfied unless 
$b=0$, contradicting the fact that $b$ should be arbitrary. Thus 
the Painlev\'e 
test is failed by this principal balance. 
 
The only way to rectify the failure of the resonance condition 
and leave $b$ as a free parameter is to modify (\ref{laurent}) by 
adding logarithm terms. More precisely taking \beq y_1\sim 
L_1(Z )+  \Delta_1(Z ), \qquad y_2\sim  L_2(Z )+ 
\Delta_2(Z ), \label{logs} \eeq the resonance condition is 
resolved by taking \beq \Delta_1\sim -\frac{8}{7} b^3 Z 
^4\log Z , \qquad \Delta_2\sim -\frac{8}{21} b^4 Z ^4\log Z 
. \label{leadlog} \eeq However, the additional terms $\Delta_1$, 
$\Delta_2$ in (\ref{logs}) must then consist of a doubly infinite 
series in powers of $Z$ and $\log Z$, with the leading order 
behaviours given by (\ref{leadlog}). Only in this way is it 
possible to represent the general solution of the system 
(\ref{sys4}) as an expansion in the neighbourhood of a singular 
point containing four arbitrary parameters. Such infinite 
logarithmic branching is a strong indicator of non-integrability 
\cite{newell, ramani}. 
 
\noindent {\bf Non-principal balance (ii):} The second balance 
denoted $(ii)$ above has resonances 
$$ 
r=-5,-1,0,6. 
$$ 
The presence of the negative integer value $r=-5$ means that this 
is a non-principal balance. (For an extensive discussion of 
negative resonances see \cite{cfp}.) This gives Laurent expansions 
\beq y_1\sim 3Z ^{-2}+ kZ ^4 -\frac{3}{2}bZ ^5 +O(Z ^7 
), \qquad y_2\sim bZ ^3 +O(Z ^5). \label{lau2} \eeq In this 
case all resonance conditions are satisfied and all higher 
coefficients in (\ref{lau2}) are determined uniquely in terms of 
$k$ and $b$. However, because it only contains three arbitrary 
constants (namely $b$, $k$ and the position $z_0$ of the 
pole), it cannot represent the general solution, but can 
correspond to a particular solution which is meromorphic. 
 
\noindent {\bf Non-principal balance (iii):} For the balance 
$(iii)$ the resonances are given by $r=-1$ and the roots of the 
cubic equation 
$$ 
r^3-15 r^2+26r+280=0, 
$$ 
which turn out to be a real irrational number and a complex 
conjugate pair, approximately 
$$ 
r=-3.2676, \quad 9.1338\pm 1.5048i. 
$$ 
While non-integer rational resonances are allowed within the weak 
extension of the Painlev\'e 
test (see \cite{weak} and section 6), 
irrational or complex 
resonances lead to infinite branching, and (as already evidenced 
by the principal balance $(i)$) the system (\ref{sys4}) cannot 
possess the Painlev\'e 
property. This non-principal 
balance may be interpreted as a particular solution corresponding 
to a degenerate limit of the general solution, and  perturbation 
of this particular solution (within the framework of the 
Conte-Fordy-Pickering perturbative Painlev\'e 
test 
\cite{cfp}) will pick up the logarithmic branching present in the 
general solution. Clearly it would have been sufficient to stop the test 
after the failure of the resonance condition in the principal 
balance $(i)$, but we wanted to present the details of the other balances 
to show the different possibilities that can arise in the singularity 
analysis of ODEs.

\section{The Ablowitz-Ramani-Segur conjecture} 

\index{Ablowitz-Ramani-Segur conjecture} 
Having considered how to test for the Painlev\'e property 
in ODEs, we now turn to the connection with 
integrable PDEs. In the 1970s it was discovered that 
ODEs of Painlev\'e type, and in particular some of the 
Painlev\'e transcendents, appeared as symmetry reductions 
of PDEs solvable by the inverse scattering technique. This led 
Ablowitz, Ramani and Segur \cite{ars} to formulate the following. 
 
\vspace{.1in} 
\noindent {\bf Ablowitz-Ramani-Segur conjecture: } 
{\it Every exact reduction of a PDE which is 
integrable (in the sense of being solvable by the inverse 
scattering transform) yields an ODE with the Painlev\'e property, 
possibly after a change of variables.} 
\vspace{.1in} 
 
\noindent To obtain ODE reductions of PDEs one can use the classical Lie 
symmetry 
method or its non-classical variants (see \cite{olver} for details), or 
the direct method of Clarkson and Kruskal \cite{clakru}. The idea is 
that having found the symmetry reductions of the PDE, one can 
either solve the ODEs that are obtained, or apply the Painlev\'e test 
to them, to see if branching occurs. If all the ODE reductions 
are of Painlev\'e type, then this suggests that the original PDE 
may be integrable. However, the need to allow for a possible 
change of variables will become apparent in section 6.  
Indeed, the most difficult aspect of this conjecture, if  
one would like to provide a proof of it, is  
in defining exactly what class of variable transformations  
should be allowed.

As an example, consider the Korteweg--de Vries (KdV) equation
\index{Korteweg--de Vries equation} 
for long waves on shallow water,  
which we write in the form 
\beq 
u_t=u_{xxx}+6uu_x. 
\label{kdv} 
\eeq 
This has three essentially different reductions to ODEs; 
details of their derivation are given in chapter 3 of \cite{olver}. 
The first 
is the travelling wave solution 
\beq 
u(x,t)=w(z), \qquad z=x-ct, 
\label{trav} 
\eeq 
where $c$ is the (arbitrary) wave speed and $w(z)$ satisfies 
\beq 
w'''+6ww'+cw'=0. 
\label{trode} 
\eeq 
After an integration and a shift in $w$ this is equivalent to 
(\ref{ellip}), and the solution of (\ref{trode}) is given by 
\beq 
w=-2\wp (z-z_0)-c/6, \label{elsol} 
\eeq 
where $z_0$ and the invariants $g_2$ and $g_3$ of the $\wp$-function are 
arbitrary constants. 
In the special case $g_2=4k^4/3$, $g_3=-8k^6/27$ the elliptic function 
degenerates to a hyperbolic function, and for $c=-4k^2$ the reduction 
(\ref{trav}) yields the one-soliton solution  
\index{Soliton}  
\beq 
\label{sol} 
u(x,t)=2k^2\mathrm{sech}^2 (kx+4k^3t). 
\eeq 
(Of course there is the additional freedom to shift the 
position of the soliton (\ref{sol}) by the transformation $x\to x-x_0$.) 
 
The second reduction of KdV is the Galilean-invariant solution 
\beq 
\label{red2} 
u(x,t)=-2\left( w(z)+t \right), \qquad z=x-6t^{2} 
\eeq 
where $w(z)$ satisfies 
\beq 
\label{2ode} 
w'''-12ww'-1=0. 
\eeq 
Upon integration, and making a shift in $z$ to remove 
the constant of integration, the ODE (\ref{2ode}) becomes 
the first Painlev\'e equation (\ref{p1}). 
 
The third reduction of (\ref{kdv}) is the scaling similarity 
solution 
\beq 
\label{red3} 
u(x,t)=(-3t)^{-\frac{2}{3}}w(z), \qquad z=(-3t)^{-\frac{1}{3}}x. 
\eeq 
This solution arises from the invariance of the PDE (\ref{kdv}) 
under the group of scaling symmetries 
$$ 
(x,t,u)\longrightarrow (\la x, \la^3 t, \la^{-2}u). 
$$ 
After substituting the similarity form (\ref{red3}) 
into KdV and integrating once we find the ODE for $w$: 
\begin{equation} 
w''+2w^{2}-zw+\frac{\ell^2-1/4+w'-(w')^{2}}{2w-z}=0. 
\label{p34} 
\end{equation} 
The parameter $\ell^2$ is the constant of integration, and 
(\ref{p34}) turns out to be equivalent to the equation P34, 
so called because it is labelled $XXXIV$ in the 
Painlev\'e classification of second order ODEs as 
detailed by Ince \cite{ince}. The equation P34 can be solved 
in terms of the second Painlev\'e equation (\ref{p2}), 
according to the relation 
\beq 
w=-y'-y^{2},    \qquad \mathrm{with}\quad \ell=\al +1/2. 
\label{mmap} 
\eeq 
The above formula defines a B\"{a}cklund  transformation 
between the two equations (\ref{p34}) and (\ref{p2}), 
and in fact there is a one-one correspondence between their solutions. 
With the parameters of the two ODEs related as in (\ref{mmap}), 
the inverse of this transformation (defined for $w\neq z/2$) is 
given by 
\[ 
y=\frac{w'+\alpha}{2w-z}. 
\] 
For more details, and higher order analogues, see \cite{nahh} 
and references. 
 
Thus we have seen that the ODE reductions of the KdV equation (\ref{kdv}) 
are solved either in elliptic functions or in terms of Painlev\'e 
transcendents, and hence these reductions certainly have the 
Painlev\'e property. So the KdV equation clearly fulfills 
the necessary condition for integrability required by the 
Ablowitz-Ramani-Segur 
conjecture,  
as it should do because it is integrable  
by means of the inverse scattering transform.  
In contrast to KdV, we consider another equation that models  
long waves 
in shallow water, namely the Benjamin-Bona-Mahoney 
(often referred to as BBM)  
equation 
\index{Benjamin-Bona-Mahoney} 
\cite{bbm}, 
which takes the form 
\beq 
u_t+u_x+uu_x-u_{xxt}=0. 
\label{bbm} 
\eeq 
The Benjamin-Bona-Mahoney 
equation is also known as the regularized long-wave equation, 
and was apparently first proposed by Peregrine \cite{peregrine}. 
The travelling wave reduction of the 
Benjamin-Bona-Mahoney equation is very similar 
to that for KdV: the PDE (\ref{bbm}) has the solution 
\beq 
u(x,t)=-12c\wp (z-z_0)+c-1, \qquad z=x-ct, 
\label{bell} 
\eeq 
given in terms of the Weierstrass $\wp$-function 
(with arbitrary values of the invariants $g_2$, $g_3$ and 
the constant $z_0$). 
In the hyperbolic limit with $c=(1-4k^2)^{-1}$ for $k\neq\pm 1/2$ 
this gives 
the solitary wave solution 
$$ 
u(x,t)=\frac{12k^2}{1-4k^2}\,\mathrm{sech}^2(kx-k(1-4k^2)^{-1}t), 
$$ 
but in contrast to (\ref{sol}) this is not a soliton because 
the Benjamin-Bona-Mahoney
equation is not integrable and collisions between 
such waves are inelastic: see the discussion and references in chapter 
10 of \cite{degm}. 
 
Evidence for the non-integrable nature of the 
Benjamin-Bona-Mahoney equation is
provided by another symmetry reduction, namely 
\beq 
u(x,t)=\frac{1}{t}w(z)-1, \qquad z=x+\ka\log t, 
\label{logred} 
\eeq 
where $\ka$ is a constant. Upon substitution of (\ref{logred}) 
into (\ref{bbm}), $w$ is found to satisfy the ODE 
\beq 
\ka w'''-w''-ww'-\ka w'+w=0. \label{logode} 
\eeq 
For all values of the parameter $\ka$, the equation (\ref{logode}) 
does not have the Painlev\'e property, which means that 
(at least in these variables) the 
Benjamin-Bona-Mahoney
equation fails the necessary condition required by the 
Ablowitz-Ramani-Segur 
conjecture. In the case $\ka =0$, (\ref{logode}) just 
becomes second order, so it is possible to compare with the 
list in chapter 14 of Ince's book \cite{ince} to see that 
$w''+ww'-w=0$ is not an ODE of Painlev\'e type. A direct method, 
which works for any $\ka$, is to apply Painlev\'e analysis directly 
to the equation and show that a resonance 
condition is failed. In fact the analysis can be greatly simplified 
by integrating in (\ref{logode}) to obtain 
\beq 
\ka w''-w'-\frac{w^2}{2}-\ka w=-\int_{z_1}^z w(s)\, ds. 
\label{intode} 
\eeq 
(The lower endpoint of integration $z_1$ is an arbitrary 
constant.) 
 
Now we can perform a Painlev\'e test on the integro-differential 
equation (\ref{intode}). For $\ka\neq 0$, in the neighbourhood of 
a movable singularity at $z=z_0$ the dominant balance is between 
the $w''$ and $w^2$ terms, giving 
$$ 
w(z)\sim 12\ka (z-z_0)^{-2}, \qquad z\to z_0. 
$$ 
If we suppose that this is the leading order in a Laurent 
expansion around $z=z_0$, i.e. 
\beq 
w(z)=\sum_{j=0}^\infty w_j (z-z_0)^{j-2}, \qquad w_0=12\ka , 
\label{laulog} 
\eeq 
then at the next order we see that the coefficient of 
$(z-z_0)^{-1}$ is 
$$ 
w_1=\frac{12\ka}{6\ka -1}, \qquad \ka\neq 1/6. 
$$ 
(For $\ka =1/6$ the Laurent expansion immediately 
breaks down.) However, 
substituting the expansion (\ref{laulog}) into the left hand side 
of (\ref{intode}) gives a Laurent series, while on the right hand 
side there is a term $\log (z-z_0)$ arising from the nonzero 
residue $w_1\neq 0$. Hence the expansion (\ref{laulog}) 
cannot satisfy the equation (\ref{intode}), or equivalently 
(\ref{logode}), and the Painlev\'e test is failed. 
 
Thus we have seen that all of the ODE reductions of the KdV 
equation possess the Painlev\'e property, but not all the 
reductions of the non-integrable 
Benjamin-Bona-Mahoney 
equation (\ref{bbm}) 
are of Painlev\'e type. We leave it as an exercise for 
the reader to 
check whether the  
Benjamin-Bona-Mahoney equation has other  reductions 
apart from (\ref{bell}) and (\ref{logred}) (for hints see 
exercise 3.2 in \cite{olver}). However, it should be 
clear from the above that a fair amount of 
work is required when analysing a PDE in the light 
of the Ablowitz-Ramani-Segur conjecture, since one must first find all 
possible reductions to ODEs and then perform Painlev\'e 
analysis on each of them separately. Finding the symmetry 
reductions can be a difficult enterprise in itself 
(see \cite{symred} for example), but in 
the next  
section we shall see how this complication can be 
avoided by using the direct method due to  
Weiss, Tabor and 
Carnevale \cite{wtc}.  
 
\section{The Weiss-Tabor-Carnevale Painlev\'e test} 

While the symmetry reductions of PDEs are clearly indicative 
of their integrability or otherwise, it is more convenient to 
analyse the singularity structure of PDEs directly. This approach 
was pioneered by Weiss, Tabor and 
Carnevale \cite{wtc} (hence it is usually referred to as 
the WTC Painlev\'e test). However, in the context 
of PDEs with $d$ independent (complex) variables $z_1,\ldots,z_d$ 
the singularities of the solution no longer occur at isolated 
points but rather on an analytic hypersurface $\sur$ of 
codimension one, defined by an equation 
\beq 
\phi ({\bf z})=0, \qquad {\bf z}=(z_1,\ldots,z_d)\in 
{\mathbb C}^d, 
\label{smfd} 
\eeq 
where $\phi$ is analytic in the neighbourhood of $\sur$. 
The hypersurface where the singularities lie is known as 
the {\it singular manifold}, and it can be used to define 
a natural extension of the Painlev\'e property for PDEs,  
which we state 
here in the form given by Ward \cite{ward}: 

\index{Painlev\'e property for partial differential equations}  
\vspace{.1in} 
\noindent {\bf Definition 4.1. The Painlev\'e property for PDEs: } 
{\it If $\sur$ is an analytic non-characteristic complex hypersurface 
in ${\mathbb C}^d$, then every solution of the PDE which is 
analytic on ${\mathbb C}^d\backslash\sur$ is meromorphic on 
${\mathbb C}^d$. } 
\vspace{.1in} 
 
With the above definition in mind, it is natural to look 
for the solutions of the PDE in the form of a Laurent-type 
expansion near $\phi ({\bf z})=0$: 
\beq 
u({\bf z})=\frac{1}{\phi ({\bf z})^\mu} 
\sum_{j=0}^\infty \al_j({\bf z})\, 
\phi ({\bf z})^{j}, 
\label{pdexp} 
\eeq 
If the PDE has the Painlev\'e property, then the leading order 
exponent $\mu$ appearing in the denominator of 
(\ref{pdexp}) 
should be a positive integer, with the expansion 
coefficients $\al_j$ being analytic near the singular manifold 
$\phi=0$, and sufficiently many of these must be 
arbitrary functions together with the arbitrary non-characteristic 
function 
$\phi$. As mentioned in \cite{kruskal} in the context of 
the self-dual Yang-Mills equations, and further explained in 
\cite{ward}, it is important to state that $\phi$ should be 
non-characteristic because (even for linear equations) the 
solutions of PDEs can have arbitrary 
singularities along characteristics. 
 
The application of 
the Weiss-Tabor-Carnevale 
test using series of the form (\ref{pdexp}) proceeds 
as for the usual Painlev\'e test for ODEs: when the series is substituted 
into the PDE, 
equations arise at each order 
in $\phi$ which determine the coefficients $\al_j$ 
succesively, except at resonant values $j=r$, where the corresponding 
$\al_r$ are required to be arbitrary (subject to 
compatibility conditions being satisfied). The 
Weiss-Tabor-Carnevale test is only 
passed if all 
resonance conditions are fulfilled 
for every possible balance in the PDE (i.e. all consistent 
choices of $\mu$). Note that, just as for ODEs, passing the test 
merely constitutes a {\it necessary} condition for the 
Painlev\'e property: a complete proof is much harder in general, 
although in the particular case of the self-dual Yang-Mills equations 
Ward \cite{ward} was able to use twistor methods to prove that 
they satisfy the requirements of Definition 4.1.

To see how the 
Weiss-Tabor-Carnevale  test works, we will indicate the first steps 
of the analysis for the example of the KdV equation (\ref{kdv}). 
In that case, there are just two independent variables $x$ and $t$, 
so $d=2$, and there is only one dominant balance where the 
degree of the singularity for the linear term 
$u_{xxx}$ matches that for the nonlinear term $uu_x$. Substituting 
an expansion of the form (\ref{pdexp}) 
into (\ref{kdv}), with ${\bf z}=(x,t)\in
{\mathbb C}^2$, it 
is clear that this gives $\mu=2$ as the only possibility, and 
for the leading order and next-to-leading order the first 
two coefficients are determined as 
\beq 
\al_0=-2\phi_x^2, \qquad \al_1=2\phi_{xx}. 
\label{tops} 
\eeq 
This means that the expansion around the 
singular  manifold for KdV can be 
written concisely as 
\beq 
u(x,t)=2(\log \phi)_{xx} +\sum_{k=0}^\infty \al_{k+2}(x,t)\,\phi(x,t)^k, 
\label{kdvsum} 
\eeq 
where it is necessary to assume $\phi_x\not\equiv 0$ so that 
$\phi$ is non-characteristic. 
 
In general, at each order $j$ 
there is a determining equation for the coefficients of 
the series given by 
\beq 
(j+1)(j-4)(j-6)\al_j=F_j[\phi_x,\phi_t,\phi_{xt}\ldots ,\al_k;\, k<j], 
\label{deters} 
\eeq 
where the functions $F_j$ depend only on 
the previous coefficients $\al_k$ for $k<j$ and their 
derivatives, as well as the various $x$ and $t$ derivatives of $\phi$. 
It is clear from (\ref{deters}) that the resonance values 
are $r=-1,4,6$, meaning that we require $\phi$, $\al_4$ 
and $\al_6$ to be arbitrary functions of $x$ and $t$. For the KdV 
equation, apart from the standard resonance at $-1$ 
corresponding to the arbitrariness of $\phi$, the 
other necessary conditions  for $r=4,6$, namely 
$F_4\equiv 0$, $F_6\equiv 0$ 
are satisfied identically, 
and so in accordance with the 
Cauchy-Kowalevski theorem these three arbitrary functions are 
the correct number to provide a local representation (\ref{kdvsum}) 
for the 
general solution of the third order PDE (\ref{kdv}). 
We leave it 
to the reader to calculate the expressions for the higher $F_j$ 
in (\ref{deters}) and verify the resonance conditions for $F_4$ 
and $F_6$; this is a standard calculation, so we omit 
further details which can be found 
in several sources, e.g. \cite{newell, wtc}. For completeness 
we note that the issue of convergence of the expansion 
(\ref{kdvsum}) for KdV has also been completely  
resolved \cite{joshikdv}. 
 
We shall return briefly to the KdV equation in the next section, 
where we discuss how series such as (\ref{pdexp}) can be truncated 
within the {\it singular manifold method}, leading to 
B\"acklund transformations and Lax pairs for integrable PDEs, 
and by further truncation to Hirota bilinear equations for the 
associated 
tau-functions. Before doing so, we would like to illustrate 
ways in which the basic Weiss-Tabor-Carnevale 
test may be further simplified, taking 
the non-integrable   
Benjamin-Bona-Mahoney equation (\ref{bbm}) as our example. 
Applying the test as outlined above directly to the equation 
(\ref{bbm}) leads to an expansion (\ref{pdexp}) very 
similar to that for KdV: it also has a single dominant 
balance with $\mu=2$ for a non-characteristic 
singular manifold (where $\phi_x\not\equiv 0\not\equiv\phi_t$), 
and the same resonances $r=-1,4,6$, but for the Benjamin-Bona-Mahoney 
equation  
not all resonance conditions are satisfied and the test is failed. 
It is a good exercise to perform this calculation and 
compare it with the corresponding results for KdV. Rather than 
presenting such a comparison here, we wish to give two shortcuts 
to the conclusion that the equation (\ref{bbm}) does not possess the 
Painlev\'e property for PDEs. First of all, observe that 
if $\phi_x\not\equiv 0$ then locally we can apply the 
implicit function theorem and solve the  equation (\ref{smfd}) for 
$x$. Thus we set 
\beq 
\phi(x,t)=x-f(t) 
\label{krud} 
\eeq 
with $\dot{f}(t):=df/dt\not\equiv 0$, and then we can take the 
coefficients in the 
expansion (\ref{pdexp}) to be functions of $t$ only; this is 
referred to as the `reduced ansatz' of Kruskal, first  
suggested in \cite{kruskal}. With this ansatz, the 
Weiss-Tabor-Carnevale analysis 
for PDEs becomes only slightly more involved than applying the 
Painlev\'e test for ODEs, and so constitutes a very effective 
way to decide if a PDE is likely to be integrable. 
 
For the Benjamin-Bona-Mahoney equation there is a second 
shortcut that can be made, 
which is to take the potential form of the equation by making use 
of the fact that it has a conservation law. 
This approach is widely applicable, since nearly all 
physically meaningful PDEs admit one or more conservation laws. 
For the equation (\ref{bbm}) it is immediately apparent that it can be 
put in conservation form as 
$$ 
\frac{\partial u}{\partial t}=\frac{\partial }{\partial x} 
\left(u_{xt}-\frac{1}{2}u^2-u\right), 
$$ 
which implies that 
$$ 
C=\int_{-\infty}^\infty u\,dx 
$$ 
is a conserved quantity for the Benjamin-Bona-Mahoney equation, i.e. 
$dC/dt=0$ for $u(x,t)$ 
defined on the whole real $x$-axis  with vanishing 
boundary conditions at $x=\pm\infty$. It follows that upon 
introducing the potential $v$ as the new dependent variable, with 
$$ 
v=\int_{-\infty}^xu\,dx\longrightarrow C\quad \mathrm{as}\quad 
x\to\infty, 
$$ 
we can replace $u$ by $v$ and its derivatives in (\ref{bbm}) 
to obtain the potential form of the PDE, namely 
\beq 
v_t-v_{xxt}+v_x+\frac{1}{2}v_x^2=0 
\label{pbbm} 
\eeq 
(where we have integrated once and applied the boundary 
conditions to eliminate the arbitrary function of $t$). 
If we now apply the Weiss-Tabor-Carnevale test to (\ref{pbbm}), 
at  the same time 
using the `reduced ansatz' (\ref{krud}), then we see that the only 
possible leading 
exponent in a Laurent-type expansion for  $v$ is $\mu =1$, giving 
\beq 
v(x,t)=\sum_{j=0}^\infty \be_j(t)\, (x-f(t))^{j-1}. 
\label{vsum} 
\eeq 
The equations for the coefficients $\be_j(t)$ at each 
order take the form 
$$ 
(j+1)(j-1)(j-6)\be_j=F_j[\dot{f},\ddot{f},\ldots,\be_k;\,k<j], 
$$ 
so the resonances are $r=-1,1,6$ which compares with $r=-1,4,6$ for 
the original equation (\ref{bbm}): clearly one of the resonances 
has shifted to a lower value by taking the equation in potential form 
(\ref{pbbm}). Upon substituting the series (\ref{vsum}) into the 
potential Benjamin-Bona-Mahoney equation, the leading order term is at order 
$\phi^{-4}$, giving the equation 
$$ 
-6\be_0\dot{f}+\frac{1}{2}\be_0^2=0. 
$$ 
Since $\be_0\not\equiv 0$, this determines the first coefficient as 
$$ 
\be_0=12\dot{f}. 
$$ 
However, at the next order $\phi^{-3}$ in the equation (\ref{pbbm}), we have 
the resonance $r=1$ with the condition 
\beq 
\label{rescon} 
-2\dot{\be}_0=0, \qquad \mathrm{whence} \qquad \ddot{f}=0. 
\eeq 
Since $f$ is supposed to be an arbitrary non-constant function of $t$, we 
see that the resonance condition (\ref{rescon}) is not satisfied, so the 
equation (\ref{pbbm}) fails the 
Weiss-Tabor-Carnevale Painlev\'e test, indicating the 
non-integrability of the Benjamin-Bona-Mahoney equation. 
However, observe what happens 
if $f$ is a linear function of $t$: then (\ref{rescon}) is 
satisfied, corresponding to the travelling wave reduction (\ref{bell}), 
which does have the Painlev\'e property. 
 
The only way to remove the restriction (\ref{rescon}) on the function 
$f$ would be to add a term $-(\dot{\be}/\dot{f})\log (x-f(t))$ to 
the expansion (\ref{vsum}). It has been observed \cite{picklog} 
that the inclusion of terms linear in $\log\phi$ for 
PDEs in potential form is not incompatible with integrability. 
However, in this case terms of all powers of $\log (x-f(t))$ 
are required to ensure a consistent expansion in the 
potential Benjamin-Bona-Mahoney equation (\ref{pbbm}) with three arbitrary 
functions $f$, $\be_1$ and $\be_6$ corresponding to the three 
resonances. 
 
For the reader who is interested in applying either the Painlev\'e test 
for ODEs,  
as described in section 2, or the Weiss-Tabor-Carnevale Painlev\'e test  
for PDEs, it is worth remarking that software implementations of these  
tests are now freely available. The web page  
${\tt www.mines.edu/fs \_ home/whereman}$ has algorithms written by  
D. Baldwin and W. Hereman, for instance.  
 
\section{Truncation techniques} 

Aside from the obvious application of the various Painlev\'e tests in 
isolating potentially integrable equations (for example, in the classification 
of integrable coupled KdV equations \cite{karasu}), their usefulness can be 
extended by the means of truncation techniques. The first of these is known as 
the {\it singular manifold method}, which was primarily developed in a series 
of 
papers by Weiss \cite{weiss}. The idea behind the method is that by truncating 
an expansion such as (\ref{pdexp}), usually at the zero order ($\phi^0$) term, 
it is possible to obtain a B\"acklund transformation  for the PDE. For 
such truncated 
expansions the singular manifold function $\phi$ is no longer arbitrary, 
but satisfies 
constraints. In the case of integrable equations 
that are solvable by the inverse 
scattering transform, 
the singular manifold method can be used to derive the associated 
Lax pair; for 
directly linearizable equations, 
such as Burger's equation or its hierarchy \cite{pickburgers}, 
the method instead leads to the correct linearization. 
Even for non-integrable PDEs, where the constraints on $\phi$ are much 
stronger, the singular manifold method can still be used to obtain 
exact solutions. 
Furthermore, for integrable PDEs the truncation approach can be 
carried further by 
cutting off the series {\it before} the zero order term, 
to yield tau-functions 
satisfying bilinear equations \cite{gtrw}. 
 
We will outline the basic truncation results 
for the KdV equation (\ref{kdv}), before 
presenting more detailed calculations for the nonlinear Schr\"odinger (NLS) 
equation. For KdV, the Laurent-type expansion 
(\ref{kdvsum}) can be consistently truncated at the zero order term to 
yield 
\begin{equation} 
u=2(\log \phi )_{xx}+\tilde{u}, \qquad \tilde{u}\equiv\al_2. 
\label{backsch} 
\end{equation} 
While substituting the full expansion (\ref{kdvsum}) into KdV gives an 
infinite set 
of equations (\ref{deters}) for $\phi$ and the $\al_j$, 
the truncated expansion gives only 
a finite number. The last of these equations 
does not involve $\phi$, and just says 
that $\tilde{u}$ is also a solution of KdV, i.e. 
$$ 
\tilde{u}_t=\tilde{u}_{xxx}+6\tilde{u}\tilde{u}_x. 
$$ 
The other equations (after some manipulation and integration) boil 
down to just 
two independent equations for $\phi$ and $\tilde{u}$, as follows: 
\beq 
\tilde{u}=k^2-\frac{(\sqrt{\phi_{x}})_{xx}}{\sqrt{\phi_{x}}}; 
\label{schroeq} 
\eeq 
\beq 
\frac{\phi_{t}}{\phi_{x}} = 6k^2+\left( 
\frac{\phi_{xxx}}{\phi_{x}}-\frac{3\phi_{xx}^{2}}{2\phi_{x}^{2}} \right). 
\label{schwarz} 
\eeq 
In the above, $k$ is a constant parameter.  The important feature to note is 
that since $u$ and $\tilde{u}$ are both solutions of (\ref{kdv}), the equation 
(\ref{backsch}) constitutes a B\"acklund transformation for KdV, provided that 
$\phi$ satisfies (\ref{schroeq}) and (\ref{schwarz}). For example, starting 
from the seed solution $\tilde{u}=0$, the B\"acklund transformation 
defined by (\ref{backsch}), 
(\ref{schroeq}) and (\ref{schwarz}) can be used to generate the one-soliton 
solution 
(\ref{sol}), or even a mixed rational-solitonic solution by taking 
$\phi= 
(x-12k^{2}t)+(2k)^{-1}\sinh (2kx+8k^{3}t)$. 
 
It is maybe not immediately obvious that the system comprised of the 
two equations (\ref{schroeq}) and (\ref{schwarz}) is equivalent to the 
standard Lax pair for KdV. This can be seen by making the squared 
eigenfunction substitution $\phi_x=\psi^2$, so that (\ref{schroeq}) 
becomes a linear (time-independent) 
Schr\"odinger equation. In the context of 
quantum mechanics in one dimension, $\psi$ is the wave function 
with potential $-\tilde{u}$ and 
energy $-k^2$, i.e. (\ref{schroeq}) is equivalent to 
$$ 
\psi_{xx}+\tilde{u}\psi=k^2\psi. 
$$ 
The second equation (\ref{schwarz}) is known as the 
Schwarzian KdV equation \cite{nhj}, 
and in its own right it constitutes 
a nonlinear integrable PDE for the dependent variable 
$\phi$; with the squared eigenfunction substitution it leads to  the 
linear equation for the time evolution $\psi_t$. All these results for 
KdV are well 
known, and have been extended to the whole KdV hierarchy; 
the interested reader 
who wishes to check these calculations 
is referred to \cite{newell} for more details. 
 
Perhaps less well understood, 
however, is the interesting connection \cite{gtrw} 
between the singularity structure of PDEs 
and the tau-function approach to soliton equations pioneered by 
Hirota \cite{hirota, nimmo}, 
which culminated in the Sato theory relating integrable systems 
to representations 
of affine Lie algebras \cite{miwa, sato}. 
The link with the singular manifold method is made 
by truncating the  expansion 
(\ref{kdvsum}) at the last singular term in $\phi$, 
and setting $\phi=\tau$, to 
give 
\begin{equation} 
u=2(\log \tau )_{xx}, \label{thkdvtau} 
\end{equation} 
which is the standard substitution for the KdV variable $u$ in terms of its 
tau-function. From (\ref{kdv}), after substituting (\ref{thkdvtau}) and 
performing an integration (subject to suitable boundary 
conditions), a bilinear equation is obtained for the new dependent 
variable $\tau$. This bilinear equation may be written concisely as 
\begin{equation} 
(D_{x}D_{t}-D_{x}^{4})\tau\cdot\tau=0, \label{thkdvbil} 
\end{equation} 
by making use of the Hirota derivatives: 
\[ 
D_{x}^{j} D_{t}^{k}g\cdot f:= 
\left(\frac{\partial}{\partial x}-\frac{\partial}{\partial x'} 
\right)^{j} 
\left(\frac{\partial}{\partial t}-\frac{\partial}{\partial t'} 
\right)^{k}g(x,t)f(x',t') 
|_{x'=x,t'=t}. 
\] 
The bilinear form is particularly convenient for calculating multi-soliton 
solutions \cite{hirota}, and leads to the connection with vertex operators 
\cite{miwa, nimmo, sato}. For solitons the tau-function is just a polynomial 
in exponentials. In general $\tau$ is holomorphic, so from (\ref{thkdvtau}) 
it is clear that the places where  $\tau$ vanishes correspond to the 
singularities 
of $u$. 
 
We now present details on the application of the singular manifold method 
to the nonlinear Schr\"odinger equation 
\index{Nonlinear Schr\"odinger equation} 
\beq 
i\psi_t+\psi_{xx}-2|\psi|^2\psi=0. \label{nls} 
\eeq 
This PDE (commonly referred to as NLS) describes the evolution of a 
complex wave 
amplitude $\psi$, and due to the minus 
sign in front of the cubic nonlinear term this is the non-focusing case of the  
nonlinear Schr\"odinger equation; 
the focusing case has $+2|\psi|^2\psi$ instead, and describes a different 
physical 
context. 
The following results on the singular manifold 
method for the nonlinear Schr\"odinger equation appeared in \cite{mythesis}. 
Seeking an expansion of the form (\ref{pdexp}) for (\ref{nls}), at 
leading order we find the behaviour 
$$ 
\psi\sim \frac{\al_0}{\phi}, \qquad |\al_0|^2=\phi_x^2. 
$$ 
Thus, truncating the expansion at the zero order  ($\phi^0$) level, we find 
\beq 
\psi=\frac{\al_0}{\phi}+\hps , \qquad \hps\equiv \al_1. \label{nlstop} 
\eeq 
To proceed with the singular manifold method we substitute the truncated 
expansion (\ref{nlstop}) into (\ref{nls}), and set the terms at each order 
in $\phi$ to zero. This yields the following four equations (the singular 
manifold 
equations): 
 
\beq 
\begin{array}{rrcc} 
\phi^{-3}:  & 
 
|\al_0|^2 - 
\phi_x^2 
 & = & 0; 
\\ 
\\ 
 \phi^{-2}:  & 
 
 i\phi_t+2\phi_x(\log \al_0 )_x+\phi_{xx} 
+2\al_0\overline{\hps} +4\overline{\al}_0\hps  & = & 0; 
\\ 
\\ 
 \phi^{-1}:   & 
 i\al_{0,t}+\al_{0,xx}-4\al_0|\hps |^2 
-2\overline{\al}_0\hps^2   & = & 0; 
\\ 
\\ 
 \phi^0:   & 
 i\hps_t+\hps_{xx}-2|\hps |^2\hps   & = & 0. 
\label{n0} 
\end{array} 
\eeq 
 
Clearly the coefficient of $\phi^{-3}$ just gives the leading order 
behaviour, while the $\phi^0$ equation in (\ref{n0}) means 
that the truncated expansion (\ref{nlstop}) constitutes an 
auto-B\"acklund transformation for  
the nonlinear Schr\"odinger equation, since $\hps$ is 
another solution of 
(\ref{nls}). Observe that for $x$ and $t$ real, 
the singular manifold function $\phi$ is seen to 
be real-valued from the leading order behaviour. Since the 
Painlev\'e analysis is really concerned with singularities in the 
space of complex $x,t$ variables, it is more consistent to 
write the nonlinear Schr\"odinger equation, together with its complex 
conjugate, 
as the system 
\beq \label{akns} 
\begin{array}{rcc} 
i\psi_t+\psi_{xx}-2\psi^2\bps & = & 0, \\ 
 
-i\bps_t+\bps_{xx}-2\bps^2\psi & = & 0, 
\end{array} 
\eeq 
and then treat $\psi$ and $\bps$ as independent quantities. 
The system (\ref{akns}) is the first non-trivial flow in the    
Ablowitz-Kaup-Newell-Segur (AKNS) hierarchy \cite{akns}. 
For this full system the singular manifold equations (\ref{n0}) should 
be augmented with the corresponding `conjugate' equations: formally 
these are obtained by taking the complex conjugate with $\phi$ real 
(as for real $x$ and $t$), and $\al_0$, $\psi$ and $\hps$ complex. 
By formally taking the real and imaginary parts of the second equation 
in  (\ref{n0}), which are equivalent to linear 
combinations of that equation together with 
its conjugate, the 
following consequences arise: 
\beq 
\begin{array}{rcc} 
\phi_{xx}+\overline{\al}_0\hps +\al_0\overline{\hps} & = & 0; 
\\ 
i\phi_t+\phi_x(\log [\al_0/\overline{\al}_0])_x 
+\overline{\al}_0\hps - \al_0\overline{\hps} & = & 0. 
\end{array} 
\label{ncons} 
\eeq 
 
Further manipulation 
of the singular manifold equations (\ref{n0}) and their conjugates, 
together with (\ref{ncons}), leads to the two equations 
\beq 
\label{finlsa} 
\al_{0,x}=-2i\la \al_0-2\hps\phi_x, 
\eeq 
\beq 
\label{finlsb} 
i\al_{0,t}=(4\la^2+2\hps\overline{\hps}) \al_0+ 
(-4i\la\hps+2\hps_x)\phi_x 
\eeq 
and their corresponding conjugates, where $\la$ is a constant. 
Upon substitution of the rearrangement 
$$ 
\al_0=(\psi -\hps )\phi 
$$ 
of (\ref{nlstop}) into (\ref{finlsa}), we 
find 
\beq 
(\psi -\hps )_x=-2i\la(\psi -\hps )-(\psi +\hps )|\psi -\hps |, 
\label{nlsbt} 
\eeq 
where we have used the first equation (\ref{n0}) 
to substitute $\phi_x=|\al_0|=|\psi -\hps |$ in the reduction 
to real $x$ and $t$. A similar equation for 
$(\psi -\hps )_t$ is obtained by eliminating $\al_0$ and 
$\phi$ from (\ref{finlsb}), and the resulting relations 
between $\psi$ and $\hps$ together with (\ref{nlsbt}) 
constitute  a B\"acklund transformation  
for the nonlinear Schr\"odinger equation in the form 
studied by Boiti and Pempinelli, taking the special 
case $\si=0$ in the formulae of \cite{bp}. Starting 
from the vacuum solution $\hps=0$, and with zero 
B\"acklund parameter $\la=0$, this BT can be 
applied repeatedly to obtain a sequence of singular rational solutions 
of the nonlinear Schr\"odinger equation, which are described in \cite{nlscrum}. 
 
The simplest 
singular rational solution 
has a single pole, which can be fixed at 
$x=0$. 
If we denote the sequence of these rational  solutions 
$\{ \psi_n \}_{n\geq 0}$, then applying the BT (\ref{nlsbt}) 
with $\la =0$ starting from the vacuum solution 
the first three are 
\beq 
\label{rats} 
\psi_0=0, \qquad \psi_1=\frac{1}{x} \qquad 
\psi_2= \frac{-2x^3+12itx+\tau_3}{x^4+\tau_3-12t^2}, 
\eeq 
with $\tau_3$ being an arbitrary constant parameter which is 
real for real $x$ and $t$. In general these rational functions 
can be written as a ratio of polynomial tau-functions 
$\psi_n=G_n/F_n$ satisfying bilinear equations (see below). 
The zeros and poles of each $\psi_n$, which are the 
roots of the polynomials $G_n$ and $F_n$ 
respectively, evolve in $t$ according to the equations 
of Calogero-Moser dynamical systems \cite{nlscrum}.

As well as leading to the B\"acklund transformation (\ref{nlsbt}) for  
the nonlinear Schr\"odinger equation, the 
singular manifold equations also yield the 
Lax pair, upon making the squared eigenfunction 
substitution 
\beq 
\al_0=-\chi_1^2, \qquad \overline{\al}_0=-\chi_2^2. 
\label{squefn} 
\eeq 
Fixing a sign we find immediately from the first equation 
(\ref{n0}) that 
$$ 
\phi_x  = \chi_1\chi_2, 
$$ 
and then putting (\ref{squefn}) into (\ref{finlsa}), 
(\ref{finlsb}) and their conjugates gives a matrix system 
for the vector $\chi=(\chi_1,\chi_2)^T$, that is  
\beq \label{zs} 
\begin{array}{ccc} 
\chi_x & = & \lu \chi , \\ 
\chi_t & = & \lv \chi, 
\end{array} 
\eeq 
with the matrices 
$$ 
\lu=\left(\begin{array}{cc} 
-i\la & \psi \\ 
\overline{\psi} & i\la 
\end{array} \right) , \qquad 
\lv=\left(\begin{array}{cc} 
-2i\la^2-i|\psi|^2 & 2\la\psi+i\psi_x \\ 
2\la\overline{\psi}-i\overline{\psi}_x & 2i\la^2+i|\psi|^2 
\end{array} \right) 
$$ 
(where we have replaced $\hps$ by $\psi$ in $\lu$, $\lv$). 
The system (\ref{zs}) is the non-focusing analogue of the Lax pair for  
the nonlinear Schr\"odinger equation  
found by Zakharov and Shabat \cite{zs}, and for $\lu$, $\lv$ 
as above the PDE (\ref{nls}) follows from the compatibility 
condition for the matrix system, which is 
the zero curvature equation 
$$ 
\lu_t-\lv_x+[\lu ,\lv ]=0. 
$$ 
For real $\la$, these matrices are elements of the Lie algebra 
$su(1,1)$, as opposed to $su(2)$ for the case of the focusing 
nonlinear Schr\"odinger equation. 
 
To obtain the Hirota bilinear form of the nonlinear Schr\"odinger equation  
we can make a further truncation in (\ref{nlstop}), 
setting $\hps =0$, $\al_0 = G$, $\phi =F$, so that (\ref{nls}) 
becomes 
$$ 
\frac{1}{F^2}((iD_t+D_x^2)G\cdot F)-\frac{G}{F^3} 
(D_x^2F\cdot F+2|G|^2)=0. 
$$ 
The two equations in brackets can be consistently decoupled to give 
the bilinear system for the two tau-functions $F$, $G$: 
\beq \label{nlsbil} 
\begin{array}{rcc} 
 (iD_t+D_x^2)G\cdot F & = & 0;  \\ 
 
D_x^2F\cdot F+2|G|^2 & = & 0. 
\end{array} 
\eeq 
It is easy to check that 
the numerators and denominators in the rational 
functions (\ref{rats}) are particular solutions of 
the system (\ref{nlsbil}). 
The bilinear form of  
the nonlinear Schr\"odinger equation was used by Hirota to derive 
compact expressions for the multi-soliton solutions \cite{hir23}. 
A further consequence of (\ref{nlsbil}) is the bilinear 
equation 
\beq 
iD_xD_tF\cdot F-2D_xG\cdot\overline{G}=i\gamma F^2, 
\label{gameq} 
\eeq 
with a constant $\gamma$. This constant can be removed by a 
gauge transformation of the tau-functions, rescaling 
both $F$ and $G$ by $\exp [\gamma xt/2]$. Eliminating $G$ 
between (\ref{gameq}) and (\ref{nlsbil}), the  
nonlinear Schr\"odinger equation 
is then rewritten as a single trilinear equation, 
expressed  as a sum of two determinants, namely 
\beq \label{tril} 
\left| 
\begin{array}{ccc} 
F & F_x & F_t \\ 
F_x & F_{xx} & F_{xt} \\ 
F_{t} & F_{xt} & F_{tt} 
\end{array} 
\right|   + 
\left| 
\begin{array}{ccc} 
F & F_x & F_{xx}  \\ 
F_x & F_{xx} & F_{3x} \\ 
F_{xx} & F_{3x} & F_{4x} 
\end{array} 
\right| =0. 
\eeq 
The tau-function solution of the trilinear equation (\ref{tril}) 
is sufficient to determine both the modulus and the argument of 
the complex amplitude $\psi$ (see \cite{nlscrum} and 
references).

From the preceding results for the KdV and nonlinear Schr\"odinger 
equations it should be 
clear that truncation methods can be extremely powerful 
in extracting information about integrable PDEs. 
There are several refinements of the singular manifold method, 
in particular those involving truncations using M\"obius-invariant 
combinations of $\phi$ and its derivatives \cite{conte, muscon}, 
and the use of two singular manifolds for PDEs with two different 
leading order behaviours \cite{twosmm}. Probably the most elegant 
and general synthesis of these extended methods is the approach 
formulated by Pickering \cite{smm}, who uses expansions in a 
modified variable satisfying a system of Riccati equations. 
Truncation methods have even been used to derive B\"acklund transformations  
for ODEs, in  particular Painlev\'e equations \cite{odesmm}. 
However, it is uncertain whether such methods can really be made 
sufficiently general in order to constitute an algorithmic 
procedure for deriving Lax pairs for integrable sytems. 
In particular, truncation methods are not directly applicable 
to integrable PDEs which exhibit movable algebraic branching in 
their solutions, which are the subject of the next section. 
 
\section{Weak Painlev\'e tests}

There are numerous examples of integrable systems which do 
not have the strong Painlev\'e property, but which satisfy the 
weaker criterion that their general solution has at worst movable algebraic 
branching. Perhaps the simplest example is to consider a Hamiltonian 
system with one degree of freedom defined by the 
Hamiltonian (total energy) 
$$ 
H=\frac{1}{2}p^2+V(q), 
$$ where the potential energy $V$ is a polyomial in $q$ of degree 
$d\geq 5$. The equations of motion (Hamilton's equations) are 
$$ 
\frac{dq}{dt}  =  p, \qquad 
\frac{dp}{dt}  =  -V'(q), 
$$ 
which are trivially integrable by a quadrature: 
\beq 
t=t_0+\int^q
\frac{dQ}{\sqrt{2(H-V(Q))}}. 
\label{quadr} 
\eeq 
If the potential energy is normalized so that the leading term of 
the polynomial is $-2q^d/(d-2)^2$, then with $q(t)$ having a singularity at 
$t=t_0$ the integral in (\ref{quadr}) gives 
$$ 
t-t_0\sim \pm \int^q \frac{(2-d)\,dQ}{2Q^{d/2}}= \pm q^{1-d/2}, \qquad 
\mathrm{as} \quad q\to\infty. 
$$ 
(for a suitable choice of branch in the square root). 
Thus at leading order we have 
\beq 
q\sim \pm (t-t_0)^{2/(2-d)}. 
\label{leadi} 
\eeq 
For both $d=2g+1$ (odd) and $d=2g+2$ (even) $q$ is determined by the 
hyperelliptic 
integral (\ref{quadr}) corresponding to an algebraic curve of genus $g$. 
When $g=1$ 
the solution is given in terms of Weierstrass or Jacobi elliptic functions, 
and 
both $q$ and $p$ are meromorphic functions of $t$. However, for a potential of 
degree 5 or more we have $g\geq 2$, and it is clear from (\ref{leadi}) that 
$q$ has an algebraic branch point at $t=t_0$, since in that case $2/(2-d)$ is 
a non-integer, negative rational number. In fact it is easy to 
verify that (\ref{leadi}) is the leading order term of an expansion in 
powers of 
$(t-t_0)^{2/(d-2)}$. Rather than being meromorphic as in the elliptic case, 
for $d\geq 5$ the function $q(t)$ is 
generically single-valued only 
on a covering of the complex $t$-plane with an  {\it infinite} number 
of sheets, and 
has an infinite number of algebraic branch points (see \cite{af}).

Clearly for potentials of degree 5 or more, this simple Hamiltonian system 
fails the basic Painlev\'e test, and yet it is certainly integrable according 
to any reasonable definition. (Indeed, any Hamiltonian system  
with one degree of freedom is integrable in the sense that Liouville's theorem  
holds.) 
In order to avoid excluding such basic integrable systems from singularity 
classification, Ramani et al. \cite{weak} proposed an extension of 
the Painlev\'e property. 
\index{Weak Painlev\' property} 
 
\vspace{.1in} 
\noindent {\bf Definition 6.1. The weak Painlev\'e property:} {\it An 
ODE has the weak Painlev\'e property if all movable singularities of 
the general solution have only a finite number of branches.} 
\vspace{.1in} 
 
There are many examples of finite-dimensional many-body Hamiltonian systems 
which are Liouville integrable and yet have algebraic branching in their 
solutions 
\cite{abenda, af}. 
Among these examples \cite{af} is the geodesic flow on an ellipsoid, which 
was solved classically by Jacobi \cite{jacobi}. 
Many other examples, such as those 
considered by Abenda and Fedorov in \cite{abenda},  arise 
naturally as stationary or travelling wave reductions of PDEs 
derived from Lax pairs, 
in particular those obtained from energy-dependent Schr\"odinger 
operators \cite{hone}. Thus the corresponding Lax-integrable PDEs 
have algebraic branching in their solutions, and fail the Weiss-Tabor-Carnevale 
test described in section 4. It is natural to extend the notion of the 
weak Painlev\'e property to PDEs as well, and perform Painlev\'e 
analysis on ODEs and PDEs with this property by allowing 
algebraic branching and rational (not necessarily integer) values 
for the resonances. We illustrate this procedure with the 
example of the Camassa-Holm equation 
and a 
related family of PDEs \cite{wanghone} 
which have peaked solitons (peakons).

The Camassa-Holm equation was derived in \cite{ch} 
by asymptotic methods as an approximation to Euler's 
equation for shallow  water waves, and was shown to be 
an integrable equation with an associated Lax pair. In 
the special case when the linear dispersion terms are removed 
the equation takes the form 
\index{Camassa-Holm equation}  
\beq 
u_t-u_{xxt}+3uu_x=2u_xu_{xx}+uu_{xxx}, 
\label{eq:caholm} 
\eeq 
and in this dispersionless limit it admits a weak solution 
known as a peakon, which has the form 
\beq 
u(x,t)=ce^{-|x-ct|}. 
\label{peak} 
\eeq 
Note that the notion of a `weak solution' (as 
defined in \cite{grindrod}, for instance) is completely 
unrelated to the `weak' Painlev\'e property. 
The peakon solution has a discontinuous derivative at the position 
of the peak, and the dispersionless Camassa-Holm equation 
(\ref{eq:caholm}) has exact solutions given by 
a superposition of an arbitrary number of 
peakons which interact and scatter elastically, just as for 
ordinary solitons. A detailed analysis of weak solutions  
of (\ref{eq:caholm}) has been performed by Li and Olver \cite{liolver}.  
 
However, the Camassa-Holm equation is an example of an 
integrable equation which does not satisfy the requirements of 
Definition 4.1, but instead passes the {\it weak} Painlev\'e 
test. In the neighbourhood of an arbitrary 
non-characteristic hypersurface 
$\phi (x,t)=0$ where the derivatives of $u$ blow up, it 
admits an expansion with algebraic branching: 
\beq \label{chexp} 
u(x,t)= -\phi_t/\phi_x+\sum_{j=0}^{\infty}\al_j(x,t)\, 
\phi^{2/3+j/3}. 
\eeq 
If we regard the branching part $\phi^{2/3}$ as the leading term 
(since it produces the singularity in the derivatives 
$u_x$, $u_t$ on $\phi=0$), then the resonances are 
$r=-1,0,2/3$ which correspond to the functions 
$\phi,\al_0,\al_2$ being arbitrary. The Camassa-Holm 
equation thus satisfies the weak extension of the Weiss-Tabor-Carnevale  
test, since the expansion (\ref{chexp}) is 
consistent, with the resonance conditions at $r=0$ and $r=2/3$ 
being satisfied.  
Of course the test is only local, whereas the 
weak Painlev\'e property is a global phenomenon, 
and to prove it rigorously 
for this PDE would require considerable further analysis. 
The weak extension of the Painlev\'e test 
is still a useful tool, in the sense that if 
an equation has irrational or complex branching 
(either at leading order or in its resonances), or 
if a failed resonance condition introduces 
logarithmic branching into the general solution, 
then this is a good indication of non-integrability. 
Nevertheless,  even for ODEs the weak Painlev\'e 
property should be applied cautiously as an 
integrability criterion. For an excellent discussion see 
\cite{ramani}.

We would now like to apply the weak Painlev\'e 
test to a one-parameter family of PDEs that includes 
(\ref{eq:caholm}), before showing the effect that changes of 
variables can have on singularity structure. We shall consider 
the family of PDEs 
\beq 
u_t-u_{xxt}+(b+1)uu_x=bu_xu_{xx}+uu_{xxx},   \label{eq:bfamily} 
\eeq 
where the parameter $b$ is constant. 
These are non-evolutionary PDEs: due to the presence of the 
$u_{xxt}$ term, (\ref{eq:bfamily}) is not an evolution equation 
for $u$. 
The (dispersionless) 
Camassa-Holm equation is the particular member of this 
family corresponding to $b=2$. The original reason for interest in this 
family is that Degasperis and Procesi applied the 
method of asymptotic integrability  \cite{dp} 
and isolated a new equation 
as satisfying the necessary conditions for integrability 
up to some order in a multiple-scales expansion. After 
removing the dispersion terms by combining a Galilean 
transformation with a shift in $u$ and rescaling, the  
Degasperis-Procesi equation can be written as 
\index{Degasperis-Procesi equation}  
\beq 
u_t-u_{xxt}+4uu_x=3u_xu_{xx}+uu_{xxx}, \label{eq:tdnodisp} 
\eeq 
which is the $b=3$ case of (\ref{eq:bfamily}), and 
it was proved in \cite{dhh} by construction of the Lax pair 
that this new equation is 
integrable.  A powerful perturbative extension 
of the symmetry approach was also applied to the 
non-evolutionary PDEs (\ref{eq:bfamily}) 
in \cite{mik}, and it was 
confirmed that only the special cases $b=2$ (Camassa-Holm) and 
$b=3$ (Degasperis-Procesi) fulfill the necessary 
conditions to be integrable. Hamiltonian structures 
and the Wahlquist-Estabrook prolongation algebra method 
for these PDEs 
have also been treated in detail 
\cite{wanghone}. Subsequently it has been 
shown that (after including dispersion) 
every member of the family  (\ref{eq:bfamily})  arises as 
a shallow water wave equation \cite{dull}, except for the special 
case $b=-1$. 
 
For Painlev\'e analysis it is convenient to 
rewrite (\ref{eq:bfamily}) in the form 
\beq 
m_t+um_x+b\,u_xm=0, \qquad m=u-u_{xx}. \label{eq:mb} 
\eeq 
To apply the weak Painlev\'e test, we look for algebraic branching 
similar to the leading order in (\ref{chexp}), with the derivatives of $u$ 
blowing up on a singular manifold $\phi (x,t)=0$. Thus we seek 
the following leading behaviour: 
\beq 
\label{umu} 
u\sim u_0+\al \phi^\mu, \qquad \mu\in\mathbb{Z}, \qquad 0<\mu <1. 
\eeq 
Then for the derivatives of $u$ and $m$ as defined 
in (\ref{eq:mb}) the most singular terms are as follows: 
$$ 
u_x\sim \al\phi_x\mu\phi^{\mu-1}, 
\qquad m\sim -\al\phi_x^2\mu (\mu-1)\phi^{\mu-2}, 
$$ 
$$ 
m_x\sim -\al\phi_x^3\mu(\mu-1)(\mu-2)\phi^{\mu-3}, 
\qquad m_t\sim -\al\phi_x^2\phi_t\mu (\mu-1)(\mu-2)\phi^{\mu-3}. 
$$ 
Substituting these leading orders into (\ref{eq:mb}) 
we find a balance at order $\phi^{\mu-3}$ between the 
$m_t$ and $um_x$ terms provided that 
$$ 
u_0=-\phi_t/\phi_x. 
$$ 
The next most singular term in the PDE is then at order 
$\phi^{2\mu-3}$, corresponding to a balance between the 
$um_x$ and $u_xm$ terms in (\ref{eq:mb}), 
with coefficient 
$$ 
-\al^2\phi_x^3\mu(\mu-1)(\mu-2 +b\mu), 
$$ 
and this is required to vanish giving 
\beq 
\label{mud} 
\mu =\frac{2}{1+b}. 
\eeq 
 
Thus 
we see that for a weak Painlev\'e expansion 
with the leading 
exponent $\mu$ 
being a rational number between zero and one, 
the most singular terms require that the parameter 
$b$ should also be rational with 
$$ 
b=\frac{2}{\mu}-1>1. 
$$ 
To find and test the resonances in an expansion 
with this leading order, it is sufficient to take the 
reduced ansatz (\ref{krud}) for $\phi$, and then 
make a perturbation of the leading order terms with 
parameter  $\ep$: 
\beq 
u\sim \dot{f}(t)+\al (t)\phi^\mu (1+\ep \phi^r), \qquad 
\phi=x-f(t). 
\label{resep} 
\eeq 
Substituting the perturbed expression into (\ref{eq:mb}) 
and keeping only terms linear in $\ep$, we see that 
terms possibly appearing at order $\phi^{\mu +r-3}$ 
cancel out automatically (due to the form of $u_0$), 
leaving the resonance equation 
coming from the coefficient of $\phi^{2\mu +r-3}$, which is 
$$ 
-\ep\al^2(r^3+(2\mu -1)r^2+2(\mu -1)r)=0. 
$$ 
Hence the resonances are 
$$ 
r=-1,\, 0,\,2(1-\mu ), 
$$ 
with $\mu$ given in terms of the parameter $b$ by (\ref{mud}). 
 
Having applied the first part of the weak Painlev\'e test and 
found a dominant balance and the corresponding values for the resonances, 
it becomes apparent that the test is completely ineffective as a means 
to isolate the two integrable cases $b=2$ and $b=3$ of 
(\ref{eq:mb}). 
Although 
the leading order resonance $r=0$ (corresponding to $\al$ being 
arbitrary) is automatically satisfied, the second resonance 
condition at $r=2(1-\mu )$ must  be checked for every rational 
value of $\mu$ with $0<\mu <1$ (or equivalently every rational value 
of the parameter $b>1$). If we write $\mu$ in its lowest terms as a ratio of 
positive integers, 
$\mu=N_1/N_2$, then (\ref{umu}) is the leading part of an 
expansion for $u$ in all powers of $\phi^{1/N_2}$, and as the difference 
$N_2-N_1$  increases there is an increasingly large number of terms to compute 
before the final resonance is reached. Checking this resonance 
for the whole countable infinity of rational numbers $b>1$ seems 
to be a totally intractable task. Gilson and Pickering showed that 
all the PDEs within a class including (\ref{eq:mb}) failed 
every one of 
a combination of {\it strong} Painlev\'e tests \cite{pick}. 
Nevertheless, it is simple to verify 
that the weak Painlev\'e test is satisfied for the two 
particular cases $b=2,3$ which are known to be integrable. 
 
However, after a judicious change of variables, involving a transformation 
of hodograph type, it is still possible to use Painlev\'e analysis 
to isolate the two integrable peakon equations. Such transformations 
have been applied to integrable PDEs with algebraic branching 
(see \cite{hodo}) in order to obtain equivalent systems with the 
strong Painlev\'e property. That this should be possible 
is in accordance with the  Ablowitz-Ramani-Segur  
conjecture, but the difficulty lies in finding the correct 
change of variables.  
In fact, for a general class of systems   
that display weak Painlev\'e behaviour (related to energy-dependent  
Schr\"odinger operators)  
we presented a particular transformation in \cite{hone}  
and, from an examination of a principal balance, we asserted 
(without proof) that   
this transformation produced equivalent systems with the strong Painlev\'e  
property. However, from a more careful calculation of other balances  
we have recently observed that this 
earlier assertion was incorrect \cite{hnv}.     
In the case of the Camassa-Holm 
equation (\ref{eq:caholm}), a link to the first negative 
flow in the KdV hierarchy was found by Fuchssteiner \cite{fuchss}, 
and in \cite{wanghone} it was shown that the appropriate transformation 
can be extended to (almost) every member 
of the family of non-evolutionary PDEs (\ref{eq:mb}).

The key to a suitable change of variables 
for (\ref{eq:mb}) is the fact that for any $b\neq 0$, 
$\int m^{1/b}\, dx$ is a conserved quantity, 
with the 
conservation law 
\beq 
p_t=-(pu)_x, \qquad m=-p^b. 
\label{eq:con3} 
\eeq 
This allows a reciprocal transformation, 
defining new independent variables 
$X,T$ via 
\beq 
dX=p\, dx-pu\,dt, \qquad dT=dt. 
\label{eq:recip} 
\eeq 
Observe that the closure condition $d^2X=0$ for the exact 
one-form $dX$ is precisely (\ref{eq:con3}), and 
transforming the derivatives yields the 
new conservation law 
\beq 
(p^{-1})_T=u_X. \label{eq:ncon} 
\eeq 
In the old variables, $p$ is related to $u$ by 
\beq 
p^b=(\partial_x^2-1)u, \label{eq:rel} 
\eeq 
Replacing $\partial_x$ by $p\partial_X$ 
and using (\ref{eq:ncon}), this means that 
(\ref{eq:rel}) can be solved for $u$ to give 
the identity 
\beq 
u=-p(\log p)_{XT}-p^b.  \label{eq:uid} 
\eeq 
Finally the conservation law (\ref{eq:ncon}) 
can be written as 
an equation for $p$ alone, by substituting back for 
$u$ as in (\ref{eq:uid}) to obtain 
\beq 
\frac{\partial}{\partial T}\left( 
\frac{1}{p}\right)+\frac{\partial}{\partial X} 
\Big(p(\log p)_{XT}+p^b\Big)=0. 
\label{eq:mbrt} 
\eeq 
 
Thus we have seen that 
for each $b\neq 0$, the equation (\ref{eq:mb}) 
is reciprocally transformed to (\ref{eq:mbrt}), with the 
new dependent variable $p$ and new independent variables 
$X$, $T$ as in (\ref{eq:recip}). (For more background 
on reciprocal transformations, see \cite{rogers1}.) 
By making the subsitution $p=\exp (i\eta)$, (\ref{eq:mbrt}) 
becomes a generalized equation of sine-Gordon type \cite{wanghone}. 
The point of making the reciprocal transformation is that 
we may now apply the {\it strong}  
Weiss-Tabor-Carnevale Painlev\'e test to the 
equation in these new variables. At leading order near a 
hypersurface $\phi (X,T)=0$ there are two types of 
singularity that can occur in the equation (\ref{eq:mbrt}), 
corresponding to 
$p$ either vanishing or blowing up there: 
\begin{itemize} 
\item $p\sim \al\phi$, for $b\geq -1$, with 
$\al =\pm \phi_X^{-1}$ for 
$b\neq -1$; 
\item $p\sim \be\phi^\mu$, for $\mu=2/(1-b)<1$. 
\end{itemize} 
In the first balance, the resonances are $r=-1,1,2$. 
However, 
if we require the strong Painlev\'e test to hold we see 
that we must have $b\in\mathbb{Z}$, since otherwise 
the $p^b$ term will introduce branching into the expansion 
in powers of $\phi$. The second balance can only hold for 
$|b|>1$, but if $b<-1$ then $\mu\not\in\mathbb{Z}$, while 
if $b>1$ then requiring $\mu =1-M$ to be a (negative) integer 
gives 
\beq 
\label{wseq} 
b=\frac{M+1}{M-1}, \qquad M=2,3,4,\ldots 
\eeq 
From the first balance we require $b$ to be an integer, 
and the  only integer values in the sequence (\ref{wseq}) are 
$b=2,3$ (corresponding to $M=3,2$ respectively). 
Interestingly, 
when the 
Wahlquist-Estabrook method is applied to 
(\ref{eq:mb}), this same sequence crops up from  
purely algebraic considerations \cite{wanghone}. 
 
The above analysis shows that the two integrable cases 
$b=2,3$ are isolated immediately just by looking at the 
leading order behaviour. It is then straightforward to show 
that for both types of singularity in the equation (\ref{eq:mbrt}), 
these two cases fulfill the resonance conditions and thus 
satisfy the strong Painlev\'e test. However, the observant 
reader will notice that further analysis is required to 
exclude the two special integer values $b=\pm 1$, for which 
only the first type of singularity arises; this is left as 
a challenge to the reader. 
 
\section{Outlook}  
 
It should be apparent from our dicussion that the 
various Painlev\'e tests are excellent heuristic 
tools for identifying whether a given system of 
differential equations is likely to be integrable or not. 
However, the strong Painlev\'e property is clearly 
too stringent a requirement, since it is not satisfied 
by a large class of integrable systems which have movable algebraic  
branch points in their solutions. On the other hand, 
checking all possible resonances in 
the weak Painlev\'e test can be impractical as a 
means to isolate integrable systems, and 
if there are negative resonances then more 
detailed analysis may be necessary to pick up 
logarithmic branching \cite{pickweakext}. 
In this short review we have concentrated 
on methods for detecting movable poles 
and branch points. However, for equations 
like (\ref{eq:caholm}), the existence of 
the peakon solution (\ref{peak}) has led to the 
promising suggestion that Dirichlet series (sums 
of exponentials) may be a useful means of testing 
PDEs \cite{pickdir}. Also, although we have only considered 
singularities of ODEs in the finite complex plane, there are extensive 
techniques for analysing asymptotic behaviour at infinity 
\cite{sachdev, tovbis, wasow}. 
 
Before closing, we should like to give a brief mention  
to the fruitful connection between  
the singularity structure and integrability of  
discrete systems, in the context of birational maps  
or difference equations. In the last twenty years, there has been  
increased interest in discrete integrable systems.  
Liouville's theorem on integrable  
Hamiltonian systems extends naturally to the setting of symplectic maps   
or more generally to Poisson maps or correspondences \cite{rag, ves1},  
and many new examples of integrable maps have been found \cite{surisbook}.   
Grammaticos, Ramani and Papageorgiou introduced a notion of  
singularity confinement for maps or difference equations \cite{grp},  
which they used very successfully as a criterion to identify   
discrete analogues of the Painlev\'e equations, and  
they proposed that it should be regarded as a  
discrete version of the Painlev\'e property.  
 
In order to illustrate singularity confinement, we  
shall consider the second order discrete equation   
\beq  
u_{n+1}(u_n)^2u_{n-1}= \alpha q^n u_n +\beta,  
\label{dp1}  
\eeq  
which is a non-autonomous version of an equation  
of the Quispel-Roberts-Thompson type \cite{qrt},  
and can be explicitly solved in elliptic  
functions in the autonomous case $q=1$ \cite{honeblms}.  
For $q\neq 1$ the equation (\ref{dp1}) can be regarded as a discrete 
analogue of  
the first Painleve equation, 
\index{Discrete Painlev\'e equations} 
because if we set  
$u_n = h^{-2}-y(nh)$, $\alpha = 4h^{-6}$, $\beta = -3h^{-8}$, 
$q=1-h^5/4$ and take the continuum limit $h\to 0$, with $z=nh$ held fixed,  
then equation (\ref{p1}) arises at leading order in $h$.  
 
The idea of singularity  
confinement is that if a singularity is reached upon iteration of a 
discrete equation  
or map,   
then it is possible to analytically continue through it. 
\index{Singularity confinement property} 
(This is by analogy with the fact  
that the solution of an ODE with the Painlev\'e property  
has a unique analytic continuation around a movable pole.)  
In the case of (\ref{dp1}), 
a singularity will be reached if one of the iterates, say $u_{N}$, 
is zero, because this means  
that the next iterate $u_{N+1}$ is not defined.  
By redefining $\alpha$ and shifting the index $n$ if necessary, we can  
take $N=1$ without loss of generality, so $u_1=0$. The vanishing of 
$u_1$ requires that at the  
previous stage $\alpha u_{0} +\beta=0$ must hold. 
Setting $u_{-1}=a$ (arbitrary)  
and $$\alpha u_{0} +\beta=\epsilon$$ 
gives $u_1\sim \alpha^2\beta^{-2}a^{-1}\epsilon \to 0$  
as $\epsilon \to 0$, and the singularity appears at  
$$u_2\sim -\beta^4 a^2 \alpha^{-3}\epsilon^{-2}.$$  
However, subsequently we have  
$u_3\sim -q^2\alpha^2\beta^{-2}a^{-1}\epsilon$, $u_4=O(1)$ and 
further iterates  
are regular in the limit $\epsilon \to 0$. In this sense, we say that the  
singularity is confined. 
 
Although the singularity confinement criterion led to the discovery   
of many new discrete integrable systems (see \cite{bilram} and references), 
it was shown by  
Hietarinta and Viallet that it is not a sufficient condition  
for integrability \cite{hv}. In fact, they found  
numerous examples of maps of the plane defined by difference equations 
of the form  
$$  
u_{n+1}+u_{n-1}=f(u_n),  
$$  
for certain rational functions $f$, 
which have confined singularities and yet whose orbit  
structure displays the characteristics of chaos. Other  
examples of singularity confinement in non-integrable maps can be found  
in \cite{singlaur}. Nevertheless, it seems that singularity confinement  
should be a necessary condition for integrability of a suitably 
restricted class  
of maps. In fact, Lafortune and Goriely have shown that for birational maps  
in $d$ dimensions, singularity confinement is a necessary condition  
for the existence of $d-1$ independent first integrals \cite{lafgor}.  
Ablowitz, Halburd and Herbst have made an alternative proposal for extending  
the Painlev\'e property to difference equations, by using 
Nevanlinna theory \cite{ahh, nevan}, 
and this has deep connections with various algebraic or arithmetic  
measures of complexity in discrete dynamics 
(see \cite{halburd, hv, rv, silvermanA} 
and  
references).

For the reader who is interested in pursuing the subject 
of Painlev\'e analysis and its applications to both 
integrable and non-integrable equations, 
a number of excellent review 
articles are to be recommended \cite{erco, jkh, newell, ramani, tabor}, 
as well 
as the proceedings volume \cite{cargese}.

\vspace{.1in} \noindent 
{\bf Acknowledgements:} Some of this article began as a tutorial 
on the Weiss-Tabor-Carnevale 
method during the summer school {\it ``What is Integrability?''} 
at the Isaac Newton Institute, Cambridge in 2001. I would like to thank 
the students for requesting the tutorial, and for being such 
an attentive audience, especially the two mature students who 
were present 
(Peter Clarkson and Martin Kruskal). 
I am also extremely grateful to the many friends and colleagues 
who have helped me to understand different aspects of all things 
Painlev\'e, in particular Harry Braden, Allan Fordy, Nalini Joshi, 
Frank Nijhoff and Andrew Pickering.

\end{document}